\documentclass{ametsoc}
\nolinenumbers

\newcommand{\pd}[2]{ \frac{\partial #1}{\partial #2} }

\newcommand{\ab}[1]{\ensuremath{\langle #1 \rangle}}

\newcommand{\ol}{\ensuremath{\overline}}

\newcommand{\mms}{\ensuremath{\mbox{ m}^2 \mbox{ s}^{-1}}}

\journal{jpo}

\bibpunct{(}{)}{;}{a}{}{,}

\title{Transport by Lagrangian Vortices in the Eastern Pacific}

    \authors{Ryan Abernathey\correspondingauthor{Ryan Abernathey, 
     Lamont Doherty Earth Observatory of Columbia University, 
     205C Oceanography, 61 Route 9W - PO Box 1000, Palisades, NY 10964-8000, US.}}

     \affiliation{Columbia University, 
     New York, New York, USA}

\email{rpa@ldeo.columbia.edu}

    \extraauthor{George Haller}
    \extraaffil{ETH, Z{\"u}rich, Switzerland}

\abstract{
Rotationally coherent Lagrangian vortices (RCLVs) are identified from satellite-derived surface geostrophic velocities in the Eastern Pacific (180$^\circ$-130$^\circ$ W) using the objective (frame-invariant) finite-time Lagrangian-coherent-structure detection method of Haller et al.~(2016) based on the Lagrangian-averaged vorticity deviation.
RCLVs are identified for 30, 90, and 270 day intervals over the entire satellite dataset, beginning in 1993.
In contrast to structures identified using Eulerian eddy-tracking methods, the RCLVs maintain material coherence over the specified time intervals, making them suitable for material transport estimates.
Statistics of RCLVs are compared to statistics of eddies identified from sea-surface height (SSH) by Chelton et al. 2011.
RCLVs and SSH eddies are found to propagate westward at similar speeds at each latitude, consistent with the Rossby wave dispersion relation.
However, RCLVs are uniformly smaller and shorter-lived than SSH eddies.
A coherent eddy diffusivity is derived to quantify the contribution of RCLVs to meridional transport; it is found that RCLVs contribute less than 1\% to net meridional dispersion and diffusion in this sector, implying that eddy transport of tracers is mostly due to incoherent motions, such as swirling and filamentation outside of the eddy cores, rather than coherent meridional translation of eddies themselves.
These findings call into question prior estimates of coherent eddy transport based on Eulerian eddy identification methods.
} 

\begin{document}

\maketitle

%
\section{Introduction}

The mesoscale (broadly 10 - 500 km) is the most energetic scale in the ocean \citep{WorthamWunsch2014}.
Phenomenologically, the mesoscale comprises a disorderly jumble of waves, vortices, fronts, and filaments, and the word mesoscale frequently appears together with the word ``eddy."
However, a survey of the literature reveals a wide range of definitions of ``eddy,'' which is used as both an adjective and a noun. The standard Eulerian statistical perspective defines ``eddy" (an adjective) simply as a fluctuation about an Eulerian time and / or spatial mean state.
The coherent structure perspective attempts to identify specific, discrete ``eddies'' (a noun) and track them through the ocean. Here we seek to clarify the relationship between Eulerian eddy fluxes and coherent structures.

Eulerian mesoscale eddy fluxes (i.e.~statistical correlations between velocity and tracer fluctuations, a.k.a.~Reynolds fluxes) play a significant role in the transport of heat, salt, momentum and other tracers through the ocean.
Because climate models generally do not resolve the mesoscale, the sub-gridscale mesoscale flux must be parameterized based on the large-scale flow properties, commonly using a diffusive closure \citep{GentEtAl1995,TreguierEtAl1997,VisbeckEtAl1997,VollmerEden2013,BachmanFoxKemper2013}.
This important problem has motivated many studies of Eulerian eddy fluxes (and associated diffusivities) in observations and eddy-resolving models \citep[e.g.][]{MorrowEtAl1992,Stammer1998,RoemmichGilson2001,JayneMarotzke2002,VolkovEtAl2008,FoxKemperEtAl2012,AbernatheyMarshall2013,KlockerAbernathey2014,AbernatheyWortham2015}.
This body of work has been largely unconcerned with coherent structures, although \citet{AbernatheyWortham2015} did note the overlap between eddy flux spectral characteristics and the lengths scales and propagation speeds of coherent mesoscale eddies.

Many different methods have been used to identify coherent structures (CSs).
These methods fall into two general categories: Eulerian\footnote{An {\em Eulerian method} for identifying coherent structures should not be confused with the {\em Eulerian} eddy flux.}
(based on instantaneous features of the velocity field) and Lagrangian (based on time-dependent water parcel trajectories).
Early Eulerian approaches used contours of the Okubo-Weiss parameter \citep{Okubo970,Weiss1991} to identify the boundaries of eddies \citep{IsernFontanetEtAl2003,IsernFontanetEtAl2004}.
More recently, closed contours of the sea-surface height anomaly (SSH) field have been employed \citep[henceforth CSS11]{CheltonEtAl2011}.
The eddy census of CSS11 has been widely adopted by the community, likely due to its open publication on the web. Other recent Eulerian CS eddy census products include \citet{DongEtAl2011} and \citet{FaghmousEtAl2015}.
While these methods differ in certain details, they are all fundamentally similar in that they use the instantaneous velocity field (or streamfunction) to identify eddies at each snapshot in time, and then track these features from one snapshot to the next.

This Eulerian approach to eddy tracking, however, suffers from several shortcomings \citep[see][for a discussion]{Haller2015,PeacockEtAl2015}.
Firstly, the structures identified in this way are not material; the Eulerian tracking algorithms associate spatially proximal features identified at neighboring time snapshots with the a single object, but these features don't necessarily represent the same fluid.
Secondly, the structures are not objective; different observers in frames translating and rotating relative to each other will identify different flow regions as coherent.
This creates a conceptual problem because material transport by eddies should be independent of the observer, as required by basic axioms of continuum mechanics.
A related issue is that OW and SSH eddies depend on arbitrary parameters or thresholds, which are routinely tuned to match expectations derived form the same methods.
Finally, and most importantly from the perspective of transport,  OW and SSHA eddies are  {\em materially incoherent} to a significant extent; under Lagrangian advection, the supposed eddy boundaries become rapidly strained and filamented, implying that water leaks significantly across the structure boundaries inferred by the OW and SSH criteria \citep{BeronVeraEtAl2013,HallerBeronVera2013}.  

Contradictions may therefore arise when such Eulerian eddy tracking methods are applied to infer material transport, as in two recent studies.
\citet{DongEtAl2014} used Eulerian eddy tracking, together with vertical structure functions of potential temperature and salinity derived statistically from ARGO profiles, to estimate the heat and salt content materially trapped inside the eddies.
By assuming no exchange with the surrounding environment for the duration of the eddy lifetime, they estimated the meridional fluxes of heat and salt on a global scale, reaching the conclusion that ``...eddy heat and salt transports are mainly due to individual eddy movements."
\citet{ZhangEtAl2014} used a similar method to estimate the eddy mass flux. They employed tracked Eulerian eddies together with vertical structure functions to estimate the potential vorticity field surrounding the eddies.
The outermost closed potential vorticity contour was assumed to constitute an impermeable material boundary for the duration of each tracked eddy, and the eddy motion was thereby translated to a mass flux.
This method estimated the westward zonal eddy mass flux in the subtropical gyre regions to be approx.~30 Sv, a surprisingly large number which is comparable to the gyre transport itself.
These approaches might seem quite appealing because they reduce the expensive problem of observing the turbulent ocean at high spatial and temporal frequency to the more tractable one of identifying and tracking a finite number of coherent eddies.
However, the work of \citet{BeronVeraEtAl2013}, \citet{HallerBeronVera2013}, and \citet{WangEtAl2015} provides evidence that these methods strongly overestimate the degree of material coherence in mesoscale eddies, calling into question the findings.

The goal of the present study is to make a more accurate estimate of material transport due to ocean mesoscale eddies using an objective (i.e.~frame independent) Lagrangian eddy detection method applied to surface velocity fields derived from satellite altimetry.
We focus on a sector in the East Pacific which has been the setting for a number of studies on Eulerian eddy fluxes \citep{RoemmichGilson2001,AbernatheyMarshall2013,KlockerAbernathey2014,AbernatheyWortham2015}.
We apply the recently introduced Rotationally Coherent Lagrangian Vortex (RCLV) methodology based on a dynamic polar decomposition of the deformation gradient developed by \citep{Haller2016}.
The key difference between our approach and Eulerian eddy detection methods is that, by numerically advecting a dense mesh of millions of Lagrangian particles, we demonstrate (rather than assume) that our identified vortices actually remain materially coherent throughout a finite time interval, as guaranteed by their mathematical construction.
Furthermore, the full Lagrangian trajectories also allow us to estimate a more broadly-defined material eddy flux due to the entire range of turbulent motions in the flow.
By comparing this full flux with the transport due to the coherent vortices, we obtain an estimate of the relative importance of material transport by coherent structures to the full turbulent transport.
We consider the two-dimensional surface geostrophic flow as observed by satellite altimetry, as this is the only large-scale velocity observation which resolves mesoscale structures, which limits our ability to probe subsurface transport.
Nevertheless, the results strongly support the conclusion that RCLVs make only a minimal contribution to meridional eddy transport.

The paper is organized as follows.
In Sec.~2, we review the RCLV definition and the concepts of Lagrangian dispersion and diffusivity.
In Sec.~3, we describe the satellite data and the numerical approach to Lagrangian particle advection.
Sec.~4 provides some case studies of Lagrangian vortices identified by our algorithm and summarizes their statistics.
In Sec.~5, we present the eddy diffusivity and the coherent eddy diffusivity.
Sec.~6 contains discussion and conclusions.

\section{Theory of Lagrangian Transport and Rotationally Coherent Vortices}

\subsection{Eulerian Eddy Flux and Lagrangian Diffusivity}

Consider a conserved two-dimensional scalar $c(x,y)$ advected by a two-dimensional velocity field $\bm{u}(x,y,t)$ where $\bm{u} = (u, v)$.
The time- and zonal-mean meridional flux of the scalar across a latitude circle in a sector of the ocean is given by $\ol{vc}$.
The overbar represents the time and zonal average:
\begin{equation}
\ol{vc} = (L_x T)^{-1} \int_{x_0}^{x_0 + L_x} \int_{t_0}^{t_0 + T} v c dx \ ,
\end{equation}
where $L_x$ is the zonal extent of the sector and $T$ is the averaging time period.
We observe that $\ol{vc}$, as any scalar flux across a designated surface, is objective, i.e.~independent of the observer.
To capture the contribution of eddies to the mean meridional flux accurately, one therefore needs an observer-independent eddy-identification scheme.

The Lagrangian dynamics of the flow are described by the kinematic equation
\begin{equation}
\pd{\bm{X}}{t} = \bm{u} \ ,
\end{equation}
where $\bm{X} = (X,Y)$ is the position vector.
We denote the initial fluid parcel positions at $t=0$ as $\bm{x}_0 = (x_0, y_0)$. 
We can use this initial position to label the fluid parcels at a later time: $\bm{X} = \bm{X}(x_0, y_0, t)$.
For homogeneous, statistically stationary turbulent flow, \citet{Taylor1921} identified the relationship between the Eulerian mean flux $\ol{vc}$ and the Lagrangian statistics as
\begin{equation}
\ol{vc} = -K_{abs} \pd{\ol{c}}{y}
\label{eq:vc}
\end{equation}
with
\begin{equation}
K_{abs} = \frac{1}{2} \pd{}{t} \ol{Y^2} \ .
\label{eq:Kabs}
\end{equation}
Here $\ol{Y^2}$ represents the mean squared Lagrangian displacement of water parcels from their initial position, i.e.~absolute dispersion; $K$, the growth rate of this absolute dispersion, represents the single particle or absolute diffusivity \citep{Lacasce2008}.
Regardless of whether the flow statistics are truly diffusive or not, eqs.~\eqref{eq:vc} and \eqref{eq:Kabs} represent the kinematic relationship between Lagrangian displacement and Eulerian flux.
The diffusivity $K_{abs}$ expresses the fundamental transport properties of the flow, independently of the background gradient $\partial \ol{c} / \partial y$. Note that $K_{abs}$ is an objective quantity by \eqref{eq:vc}, since both $\ol{vc}$ and $\partial \ol{c} / \partial y$ are objective. (The frame-independence of $\partial \ol{c} / \partial y$ can be seen by noting that it is the directional derivative of $\ol{c}$ along the normal of a latitude circle.) Equivalently, \eqref{eq:Kabs} is objective, since the displacement $Y$ is defined relative the the reference latitude.

From an Eulerian perspective, the eddy component of the flux is readily identified via a standard Eulerian Reynolds decomposition: $\ol{vc} = \ol{v} \ol{c} + \ol{v'c'}$, where the prime indicates the instantaneous deviation from the Eulerian mean.
The second term $\ol{v'c'}$ is commonly termed the eddy flux.
Taylor envisioned a homogeneous, isotropic turbulent flow with no mean component, i.e.~$\ol{v}=0$, such that $\ol{vc} = \ol{v'c'}$.
In contrast, most geophysical flows have mean flows, and the mean advection can influence $K_{abs}$.
One way to remove the effects of the mean flow in the Lagrangian frame is to instead focus on the relative diffusivity \citep{Batchelor1952,Bennett1984}
\begin{equation}
K_{rel} = \frac{1}{2} \pd{}{t} \ol{\left( Y - \ol{Y} \right )^2} \ ,
\label{eq:Krel}
\end{equation}
which represents the growth rate of the second moment of the ensemble displacement.
In this case, the ensemble consists of all water parcels originating at a particular latitude.
(Relative diffusivity can equivalently be calculated from pair separation statistics [\citealp{Lacasce2008}].)
A detailed discussion of the relationship between $K_{abs}$, $K_{rel}$, and the mixing of a passive tracer is given by \citet{KlockerEtAl2012b}.
For the purposes of this study, we shall take $K_{rel}$ to be the most relevant diagnostic of net meridional eddy transport in our sector.
Our goal is to identify the contribution of coherent Lagrangian eddies to $K_{rel}$.
Like $K_{abs}$, $K_{rel}$ is an objective quantity.
As noted above, a self-consistent and accurate assesment of the coherent-eddy component $K_{rel}$ should also be based on objective eddy identification schemes, such as the one described next.

\subsection{Rotationally Coherent Lagrangian Vortices}

In order to partition the transport defined in \eqref{eq:Kabs} and \eqref{eq:Krel} into a contribution from coherent Lagrangian eddies, the domain must be divided into regions inside and outside a suitably defined eddy boundary.
For the boundary to be relevant for transport, it must be a material line (in 2D) or surface (in 3D) derived from an objective (frame-invariant) method.
The identification of such boundaries in unsteady turbulent flows is the subject of much recent work from the field of dynamical systems, and several possible criteria exist \citep[for a review see][]{Haller2015}.
We emphasize again that the Eulerian eddy identification methods of CSS11 are not objective and depend on choices of thresholds and parameters. Consequently, they yield boundaries which, when advected as material lines, rapidly deform and disperse away from the supposed eddy center \citep{BeronVeraEtAl2013,HallerBeronVera2013}.

One sensible criterion is to define eddy boundaries as closed material curves which experience minimal tangential stretching over a finite-time interval, so-called elliptic LCSs \citep{HallerBeronVera2012}.
A more general approach locates material eddy boundaries that exhibit uniform stretching and hence show no filamentation \citep{HallerBeronVera2013}.
The elliptic LCS detection methods have been applied to study Agulhas rings \citep{BeronVeraEtAl2013,HallerBeronVera2013,WangEtAl2015}.
The underlying variational principles guarantee prefect lack of filamentation for the boundaries and hence tend to be stringent and computationally complex.
(Recent work by \citealt{SerraHaller2017} has, however, simplified the necessary computations considerably.)

Here we opt for a fluid-mechanically more intuitive approach based on vorticity.
\citet{HallerEtAl2016} showed that rotationally coherent Lagrangian vortex (RCLV) boundaries can be identified as the outermost closed contours of the Lagrangian-averaged vorticity deviation (LAVD, defined below).
The physical essence of an RCLV is the notion that all fluid parcels along a coherent material vortex boundary should rotate at the same average angular velocity over a finite-time interval, in analogy to solid body rotation.
The LAVD technique enables the identification of such coherently rotating structure boundaries from Lagrangian trajectory data.
\citet{HallerEtAl2016} further showed that the RCLVs identified in this way coincided with structures identified by the earlier elliptic LCS methods, although the RCLV boundaries were larger\footnote
{These larger boundaries are no longer guaranteed to be completely free from filamentation under material advection.
However, by construction, any filamentation they might exhibit is trangential to the boundary, and hence the stretched boundary keeps traveling with the eddy without global breakaway.}.
Given the relative computational simplicity and the familiarity of vorticity to most physical oceanographers, we adopt this approach as our eddy identification technique.
Here we briefly review the practical elements of the theory and refer the reader to \citet{HallerEtAl2016} for a deeper mathematical treatment. 

The instantaneous relative vorticity in two dimensions is
\begin{equation}
\zeta(x,y,t) = -\pd{u}{y} + \pd{v}{x} \ .
\end{equation}
The vorticity deviation is obtained by subtracting the spatial average, i.e. $\zeta'(x,y,t) = \zeta - \ab{\zeta}(t)$.
(Angle brackets indicate an average over the whole computational domain.)
Subtracting the mean vorticity field removes any solid body rotation of the entire domain and is required to maintain the frame invariance of the method \citep{HallerEtAl2016}.
In practice, however, when the domain is the entire ocean, the mean vorticity is rather negligible.
A Lagrangian-averaged quantity is the instantaneous quantity averaged along the evolving flow trajectory (as opposed to an Eulerian average at a fixed location).
The LAVD is hence given by
\begin{equation}
LAVD_{t_0}^{t_1}(x_0,y_0) = \frac{1}{t_1 - t_0} \int_{t_0}^{t_1} | \zeta'[X(x_0,y_0,t), Y(x_0,y_0,t), t] | dt \ .
\label{eq:lavd}
\end{equation}
The LAVD is a function of position $(x_0, y_0$) but also depends on the time interval $t_0, t_1$.
RCLV boundaries at time $t_0$ are then defined as the outermost convex and closed LAVD curves surrounding local maxima of the LAVD field.
The maxima themselves are the Lagrangian vortex centers, which can be proven to be attractors or repellors for floating debris, depending on the polarity of the eddy \citep[c.f.][]{HallerEtAl2016}.
Details of the numerical computation of RCLVs are given in Sec.~4.

Once the RCLVs are identified for a specific time interval, it is straightforward to compute the associated contribution to dispersion and diffusivity.
We define a masking function $m_{t_0}^{t_1}$ to be $1$ inside each RCLV boundary and $0$ outside. The {\em coherent relative diffusivity} is then defined as
\begin{equation}
K_{rel}^{cs} =  \frac{1}{2} \pd{}{t} \ol{m_{t_0}^{t_1} \left( Y - \ol{Y} \right )^2} \ .
\label{eq:Krel_frac}
\end{equation}
The $m_{t_0}^{t_1}$ factor masks all regions that are not within a coherent structure, effectively assuming such regions move only with the mean flow and induce no relative dispersion.
By comparing $K_{rel}^{cs}$ with $K_{rel}$, we thereby quantify the fraction of meridonal eddy transport due to coherent structures.
If it is true that
eddy transport is ``mainly due to individual eddy movements" \citep{DongEtAl2014}, then $K_{rel}^{cs} \simeq K_{rel}$.
In contrast, if most of the transport is due to {\em incoherent} motion outside of the structures, then $K_{rel}^{cs} \ll K_{rel}$.
Note that $K_{rel}^{cs}$ includes two distinct modes of dispersion: coherent meridional motion of the whole eddy and rotation of the water parcels within the eddy.

\section{Satellite Data and Particle Advection}

To identify RCLVs and compute relative dispersion, we use satellite-derived surface geostrophic velocities to numerically advect virtual Lagrangian particles.
In this study, we consider only transport by the two-dimensional near-surface geostrophic velocity.
This is of course an incomplete representation of the full flow field, but the geostrophic flow is by far the dominant component at the scales of interest here.
It was shown by \citet{RypinaEtAl2012} that Ekman currents, the main large-scale ageostrophic motion in the open ocean, make a negligible contribution to mesoscale dispersion compared to the geostrophic flow.
In the conclusions, we speculate about the possible role of ageostrophic and / or unresolved motion for the detection of RCLVs.

\subsection{AVISO Surface Geostrophic Velocities}

The surface geostrophic velocity field ($\bm{v}_g$) is related to the sea surface height (SSH) relative to the geoid ($\eta$) via
\begin{equation}
\hat{\bm{k}} \times \bm{v}_g = -\frac{g}{f} \nabla \eta 
\end{equation}
where $f$ is the Coriolis parameter, $g$ is the gravitational acceleration, and $\hat{\bm{k}}$ is the unit vector pointing out of the sea surface. Satellite altimetry measures the SSH $\eta$.

We employ pre-computed gridded geostrophic velocities from the AVISO. The AVISO gridding process uses objective interpolation \citep[their terminology is unrelated to the objectivity of coherent structures]{Barnes1964} to map along-track satellite radar altimetry from various platforms onto a 1/4$^\circ$ latitude-longitude grid \citep{DucetEtAl2000}.
In addition to providing pre-computed geostrophic velocities, this product also applies a higher-order vorticity balance to estimate velocities in the equatorial region (within $\pm 5^\circ$) where geostrophy does not hold \citep{LagerloefEtAl1999}.
While this provides a complete global surface velocity field, the results in this equatorial band are less reliable.
The data were downloaded in 2015 and reflect the most recent AVISO algorithm and processing available at that time.
We use the delayed-time, reference, all-satellite merged product.
We consider the time period from Jan.~1, 1993 - Oct.~17, 2014. 

An additional processing step was undertaken: we applied a small correction to the AVISO geostrophic velocity field to remove the divergence due to the meridional variation of $f$ and to enforce no-normal-flow boundary conditions at the coastlines.
The resulting velocity field, henceforth denoted simply $\bm{v}$, is an exactly two-dimensional non-divergent flow.
The correction procedure is described in detail by \citet{AbernatheyMarshall2013}, who demonstrate that the corrections are small in magnitude compared to the original geostrophic velocity.
We compared corrected vs.~non-corrected LAVD fields and found a negligible impact on the identification of RCLVs in the open ocean.

\subsection{Advection of Lagrangian Particles}

As noted by \citet{HallerEtAl2016}, the $LAVD$ field may contain structure on smaller scales than the scales of the velocity field itself.
This is related to the fact that a relatively coarse chaotic advection field can produce very fine structure in passive tracers \citep{Pierrehumbert1991}.
Practically, it means that an extremely dense mesh of Lagrangian particles is required to properly resolve the $LAVD$.
This leads to a significant computational burden if, as here, one wishes to study a large geographical area and temporal extent.
The AVISO product is gridded at 1/4$^\circ$ resolution and resolves SSH anomalies of roughly 50 km and larger (CSS11).
However, sensitivity tests indicated that an initial particle spacing of 1/32$^\circ$ is necessary to achieve sufficient accuracy in the LAVD field and identification of RCLVs.
The mesh of initial positions is located between 180$^\circ$ and 130$^\circ$ W longitude and 80$^\circ$ S and 80$^\circ$ N latitude, a total of 8192000 points.
(In retrospect, many of the high-latitude particles were not useful, since they lie within land points or within the marginal sea-ice zone where AVISO velocities are not available.
We restrict the analysis to the latitude range 65$^\circ$ S - 60$^\circ$ N.)

The initial particle positions (at time $t_0$) determine the discrete coordinates of the $LAVD$ field.
These initial positions can therefore be chosen to facilitate the identification of RCLVs.
In particular, the first step in the algorithm requires the maxima of $LAVD$ to be identified \citep{HallerEtAl2016}.
The most obvious initial deployment, a rectangular grid, is actually not ideal for the robust identification of maxima because the relationship between diagonally connected points is ambiguous;
a better choice is a hexagonal grid, in which each points has six unambiguous neighbors \citep{Kuijper2004}.
To transform a rectangular mesh to a hexagonal one, every other row is offset by $\Delta x / 2$, where $\Delta x = 1/32^\circ$ is the spacing in the zonal direction. 

We seek to identify RCLVS with lifetimes of 30, 90, and 270 days.
Accordingly, we segment the time domain into non-overlapping $N$-day intervals (a total of 265 30-day, 88 90-day, and 29 270-day intervals).
While the specific interval bounds are arbitrary, RCLVs are structurally stable by construction, i.e.~small changes in the extraction interval will have a small effect.
If we identify a RCLV over a given time interval, this structure will generally be a subset of a larger RCLV that we would obtain in the same location for shorter time intervals.
So we would not lose any of the RCLVs if we picked shorter time intervals (unless we pick such short intervals that Lagrangian coherence can no longer be established from the available data).
Rotational coherence is a finite-time notion and hence the same water mass may become incoherent over longer times.
It is still possible that we miss some short-lived RCLVs over longer time intervals.
      
The Lagrangian trajectories are determined by solving the equation $d\bm{X}/dt = \bm{v}$ numerically using the MITgcm \citep{AdcroftEtAl2014}, an ocean general circulation model. Although this model is primarily designed for prognostic ocean simulations, it has several features that make it an attractive choice for computationally demanding Lagrangian simulations. First, it can operate in offline mode, in which velocity fields are read from files. Second, it supports Lagrangian particle tracking (via the {\tt flt} package) and implements fourth-order Runge-Kutta integration. Finally, MITgcm can run efficiently in a massively parallel configuration on many nodes of a high-performance computing cluster, providing the necessary memory and CPU performance to enable large Lagrangian ensembles.

An MITgcm run was performed for each of the temporal segments described above, and particle data was output daily. (Relative vorticity was calculated on the Eulerian grid and interpolated bilinearlly to particle positions.) The total data volume of output generated for the study was over 2 TB. The identification of RCLVs from this data is described in the following section.

\section{Identification and Statistics of Lagrangian Vortices}

\subsection{Algorithm}

The algorithm employed for identifying RCLVs follows \citet{HallerEtAl2016}, with some slight modifications for computational efficiency.
\begin{enumerate}
\item At time $t_0$, initialize a hexagonal mesh of Lagrangian particles over the domain.
\item Advect particles forward until time $t_0 + T$, where $T$ is the desired vortex lifetime (here 30, 90 and 270 days). Output particle position and relative vorticity every day.
\item Average the vorticity deviation (absolute value of relative vorticity minus global mean vorticity) over particle trajectories and map back to initial positions $\bm{x_0}$. The resulting field is the $LAVD$.
\item Identify maxima of $LADV$. (Maxima are unambiguously identified on a hexagonal mesh.) These are the RCLV centers.
\item Find the largest closed and convex curves around $LAVD$ maxima. These are the RCLV boundaries. The regions are grown iteratively by adding points; iteration stops when the next point to be added lies within the convex hull of the current region. This method admits a small convexity deficiency (usually of order 0.01) in the curves to account for the discrete nature of the numerically computed LAVD field. 
\item Filter the RCLVs by discarding features with area below a minimal admissible size (here chosen to be the area of a circle with a diameter of 30 km)
\end{enumerate}

\begin{figure}[h]
 \centerline{\includegraphics{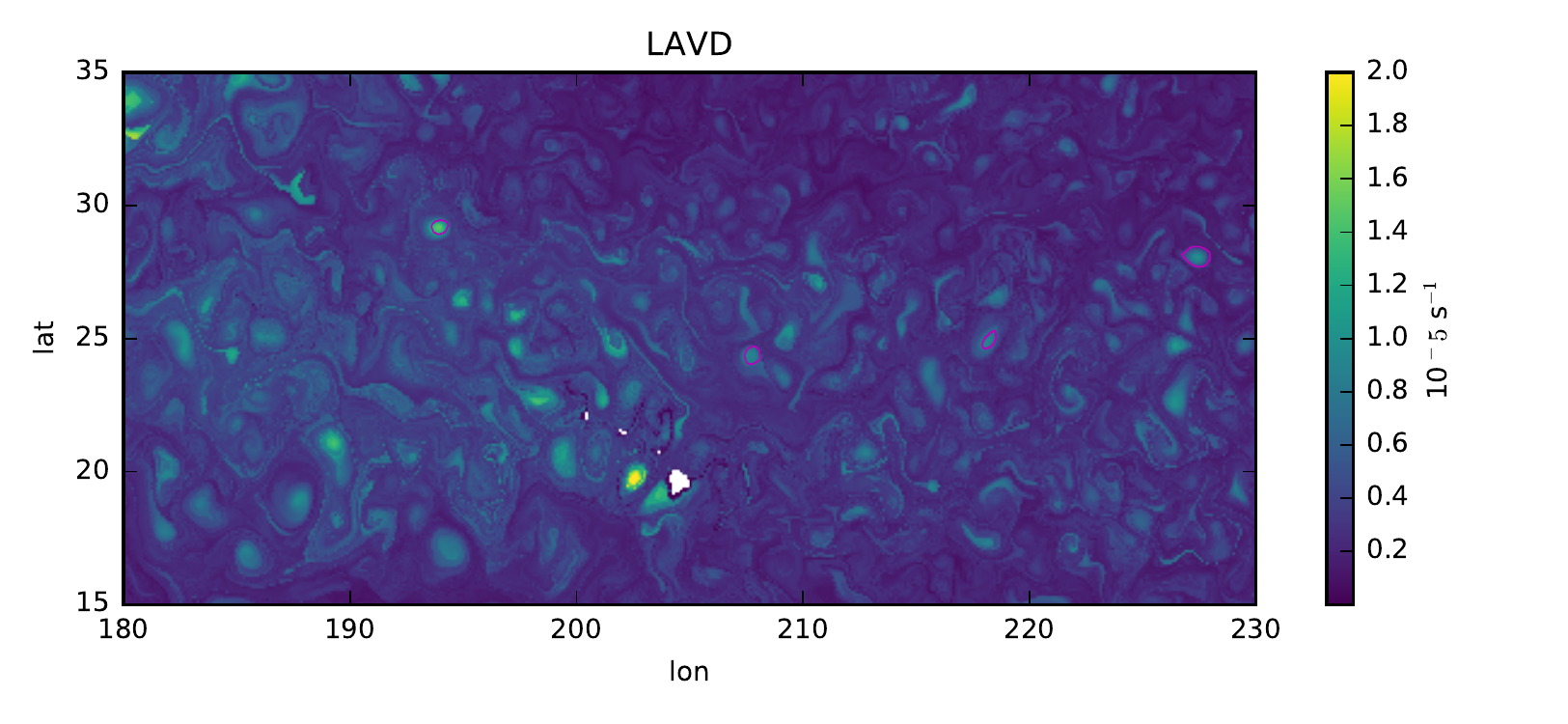}}
  \caption{A close-up example of a 90-day LAVD field from the region near Hawaii. Identified RCLV boundaries are shown as magenta contours. Note that the large local LAVD maximum just west of Hawaii is not associated with an RCLV because the LAVD field near the maximum exhibits a spiraling filamentary structure. (This is indicative of vortex breakup during the interval.)
  }\label{fig:LAVD_map}
\end{figure}

This algorithm, as well as other general-purpose data processing routines for MITgcm particle trajectories, was implemented in a python package called {\tt floater}, available at \url{https://github.com/rabernat/floater}.
An example of the $LAVD$ field, together with the positions of identified RCLVs, is shown in Fig.~\ref{fig:LAVD_map}.

\subsection{Example Vortices}

The initial and final locations of two randomly selected 90-day RCLVs are shown in Fig.~\ref{fig:example_RCLV}, superimposed on the SSH anomaly field.
One is located in the North Pacific subtropical gyre, while the other is in the Antarctic Circumpolar Current.
These vortices clearly remain materially coherent over the 90-day lifetime, consistent with the results of \citep{HallerEtAl2016}.
We examined hundreds of examples and found similar behavior.
This should be contrasted to the behavior of SSH eddies, whose boundaries are rapidly deformed under advection by the surface geostrophic flow \citep{BeronVeraEtAl2013}.

One noteworthy feature of these example RCLVs is that they do appear embedded within SSH anomalies. However, the coherent core is much smaller than the SSH anomaly.
Here we do not attempt to comprehensively compare the RCLVs with tracked SSH eddies on a feature-by-feature basis, but a statistical comparison (next subsection) suggests that this difference in size holds in general.

\subsection{Vortex Statistics}

\begin{figure}[h]
 \centerline{\includegraphics[width=\textwidth]{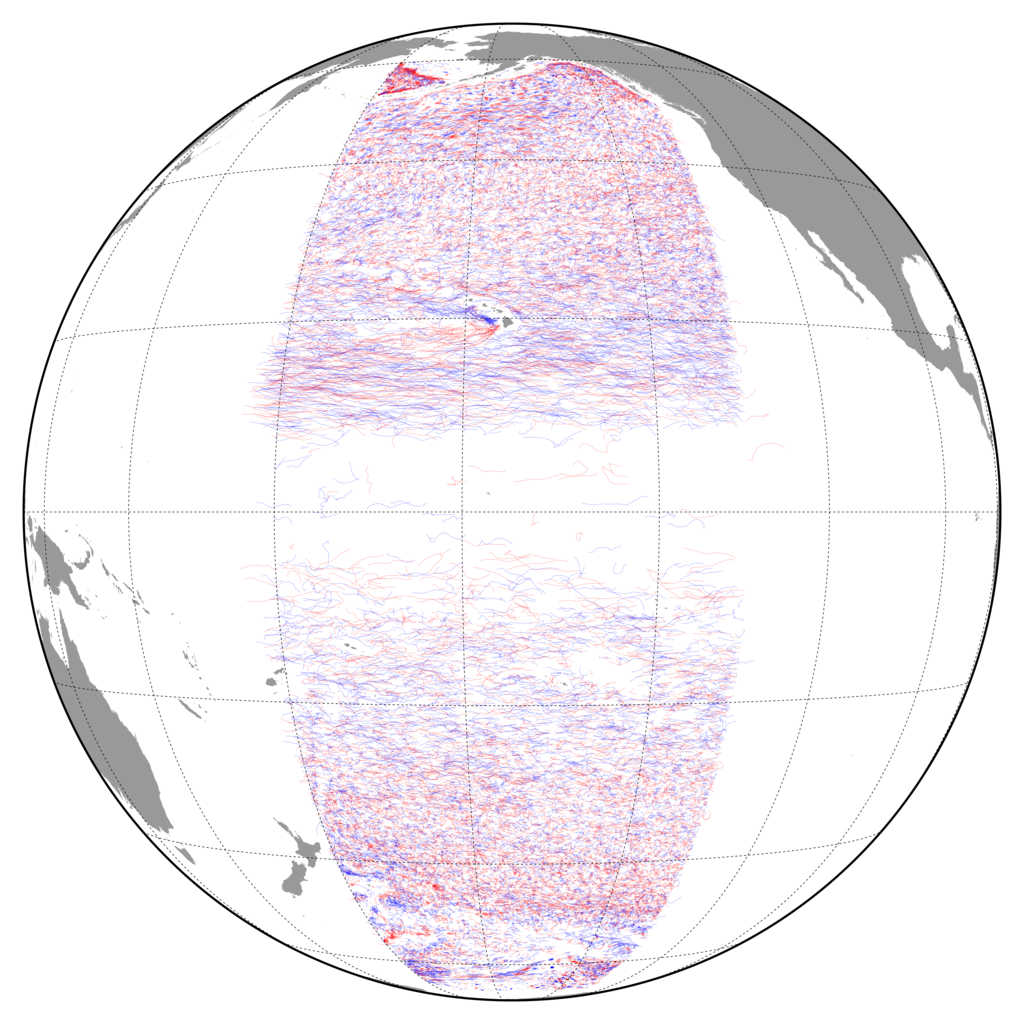}}
  \caption{The trajectories of the center points of all 46486 30-day RCLVs identified in the period 1993-2015. Trajectories represent actual Lagrangian water parcel paths under advection by the surface geostrophic flow.
  Cyclonic vorticies are shown in blue and anticyclonic are shown in red.  
  }\label{fig:trajectory_map}
\end{figure}

\begin{figure}[h]
 \centerline{\includegraphics[width=\textwidth]{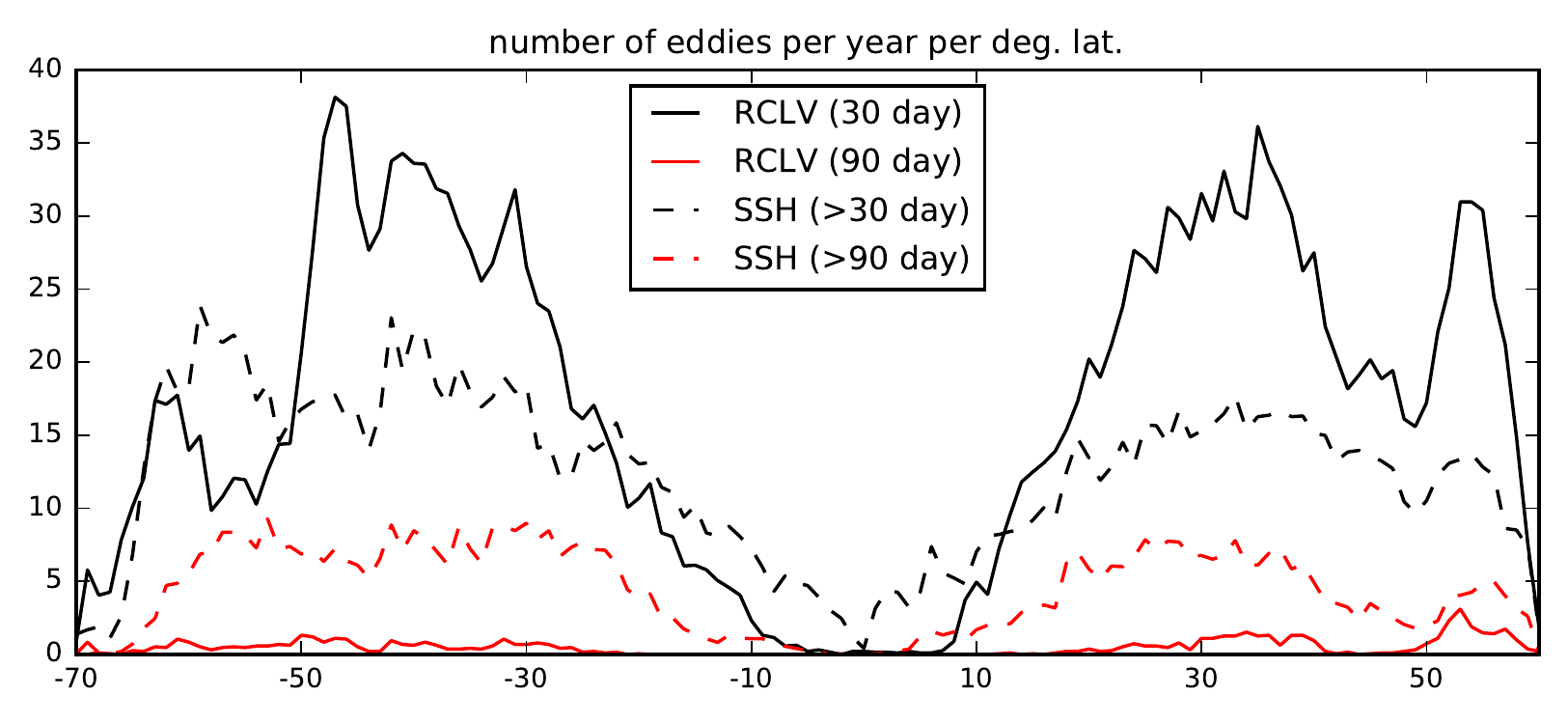}}
  \caption{Number of eddies per year per degree latitude in the east Pacific sector for RCLVs (this study; solid line) and SSH eddies \citep[][; dashed line]{CheltonEtAl2011}.
  The colors correspond to the eddy lifetime; black shows lifetimes of $\geq$ 30 days and red shows lifetimes of $\geq$ 90 days.
  }\label{fig:eddy_count}
\end{figure}

In this section we calculate some statistics of all the identified RCLVs and compare them to the statistics of the tracked SSH eddies of CSS11, whose data is publicly available at \url{http://wombat.coas.oregonstate.edu/eddies/}.
We identified 41875, 30-day RCLVs, 1182 90-day RCLVs, and only 1 270-day RCLV in the period 1993-2015.
The trajectories for all the 30-day RCLVs are plotted in Fig.~\ref{fig:trajectory_map}.
The average number of RCLVs per degree of latitude per year is plotted in Fig.~\ref{fig:eddy_count}.
(270-day RCLVs are excluded from all subsequent discussion, since they are almost non existent.)
Both figures reveal the densest concentration of RCLVs in midlatitudes, with very few RCLVs detected in the tropics. This is broadly similar to SSH eddies.

The comparison of the occurrence of 30-day vs.~90-day RCLVs in Fig.~\ref{fig:eddy_count} reveals that the shorter-lived vortices are much more prevalent.
There are actually more 30-day RCLVS that SSH eddies with lifetimes larger than 30 days.
However, the reverse is true for 90-day RCLVs; there are many more SSH eddies with equivalent or longer lifetimes.
The comparison with 270-day RCLVs, which are essentially non-existent, is even more extreme; the census of CSS11 identifies 2076 SSH eddies with lifetimes at least that long.

\begin{figure}[h]
 \centerline{\includegraphics{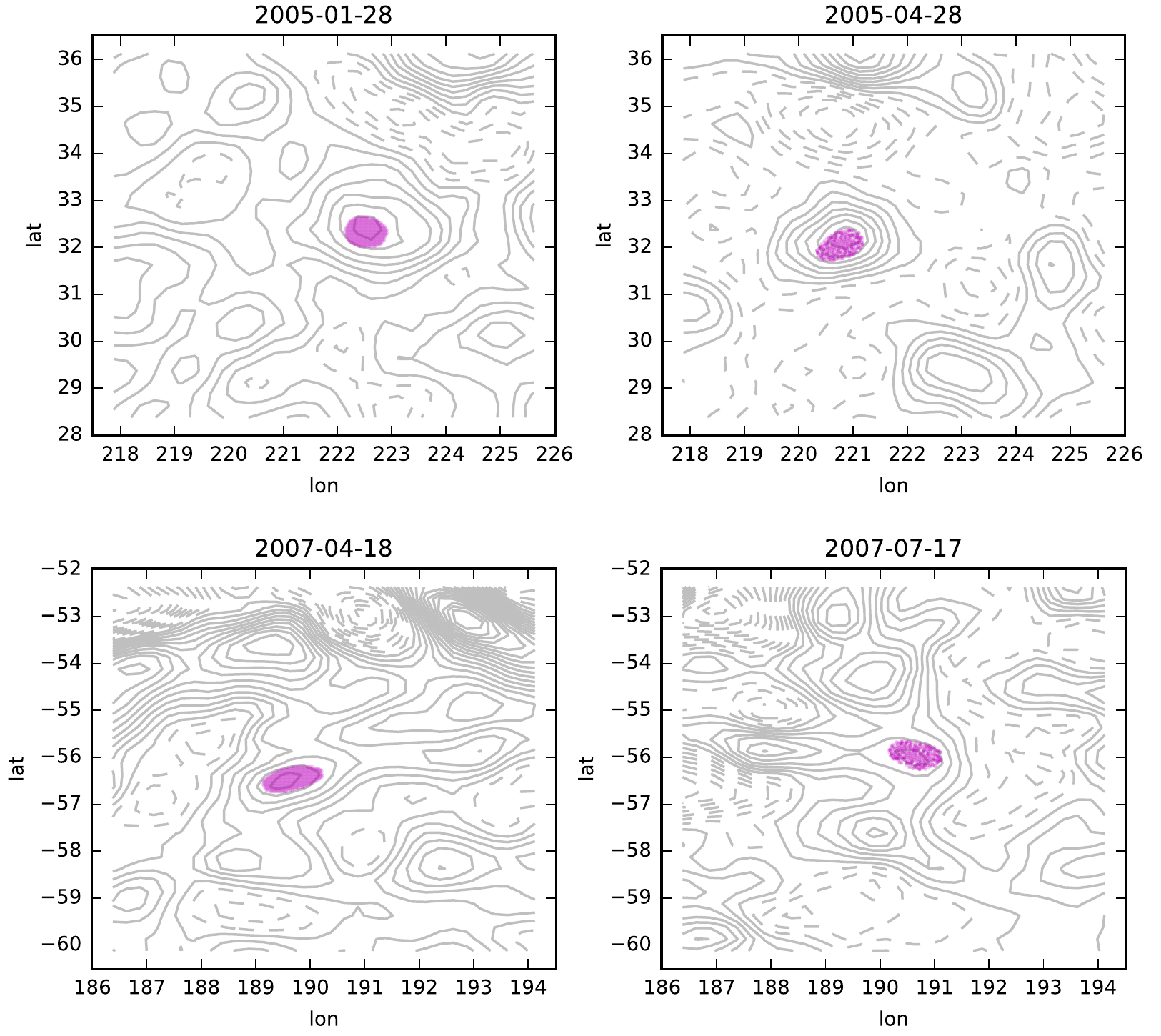}}
  \caption{The initial (left) and final (right) locations of two randomly selected 90-day RCLVs.
  The points contained within each RCLV boundary are visualized as magenta dots, overlaid on the contours
  of SSH anomaly.
  The contour interval for SSH is 2 cm. 
  }\label{fig:example_RCLV}
\end{figure}

\begin{figure}[h]
 \centerline{\includegraphics[width=\textwidth]{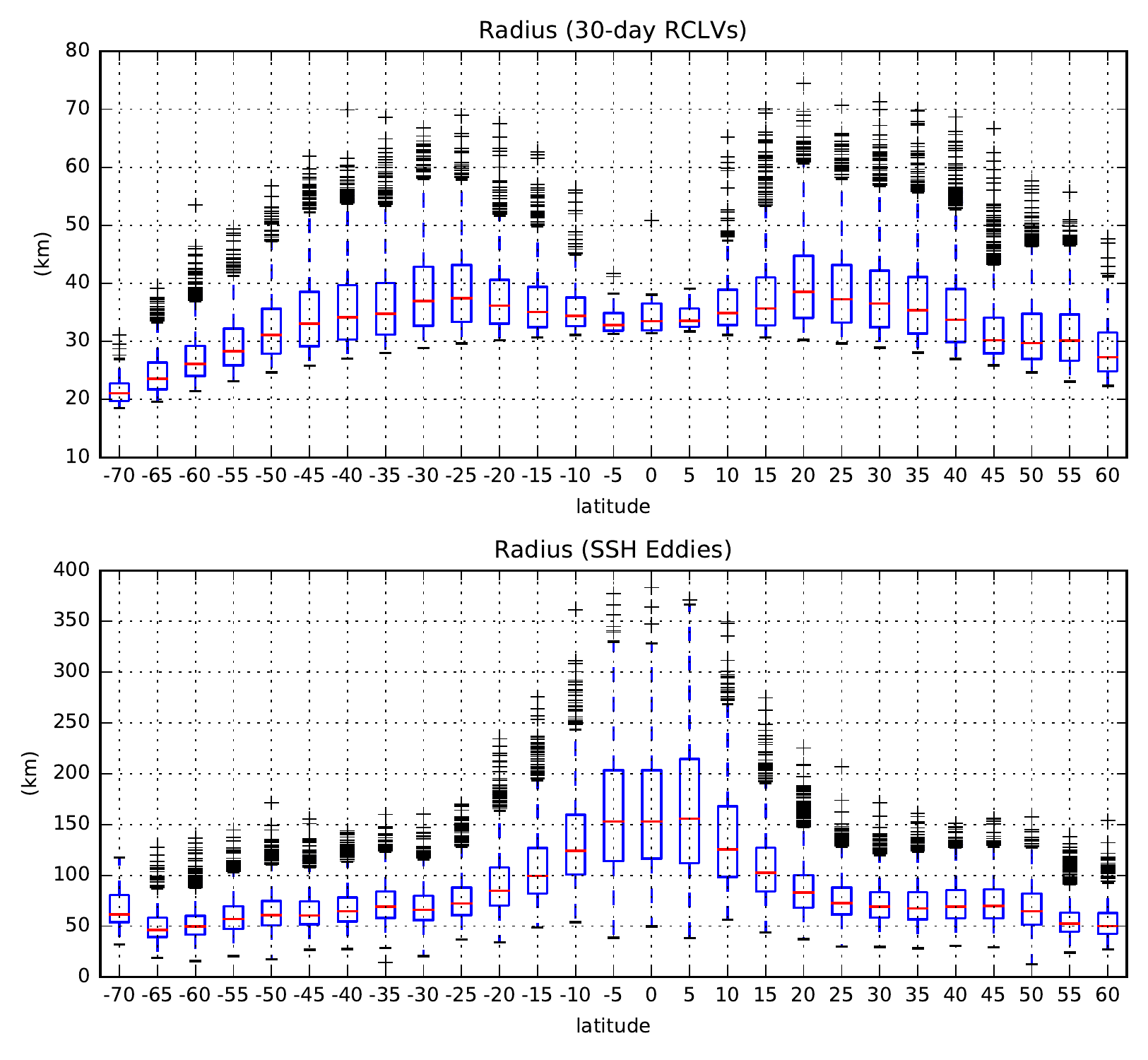}}
  \caption{Radius statistics of 30-day RCLVs (upper) and all SSH eddies (lower).
  Statistics of all eddies in 5-degree bins are shown using a box-and-whisker plot.
  The red line indicates the median. The blue box spans the middle two quartiles (25th--75th percentiles) of the distribution.
  The black whiskers span the 10th--90th percentiles. Outliers are shown using the black $+$ symbol.
  }\label{fig:radius}
\end{figure}

We now examine the statistics of eddy size. The RCLV area $A$ is converted to a radius $r$ via the formula $r = \sqrt{A/\pi}$.
Most of the RCLVs are approximately circular (e.g.~Fig.~\ref{fig:example_RCLV}), and this conversion yields a familiar unit for assessing length scales.
In Fig.~\ref{fig:radius}, we plot this radius and compare it to the radius of SSH eddies from CSS11.
Statistics are calculated in 5-degree latitude bins.

Outside of the tropics (where RCLVs are rare),  Fig.~\ref{fig:radius} reveals a familiar inverse relationship between eddy size and latitude, which likely reflects the dependence of the Rossby deformation radius on the Coriolis parameter \citep{CheltonEtAl1998}.
The largest median RCLV radius occurs near $\pm 30^\circ$, where it approaches 40 km.
Comparing with CSS11, the median RCLV radius is roughly about half of the median SSH eddy.
This is consistent with the example vortices shown in Fig.~\ref{fig:example_RCLV}.
The SSH size statistics diverge qualitatively from the RCLVs in the tropics.

\begin{figure}[h]
 \centerline{\includegraphics[width=\textwidth]{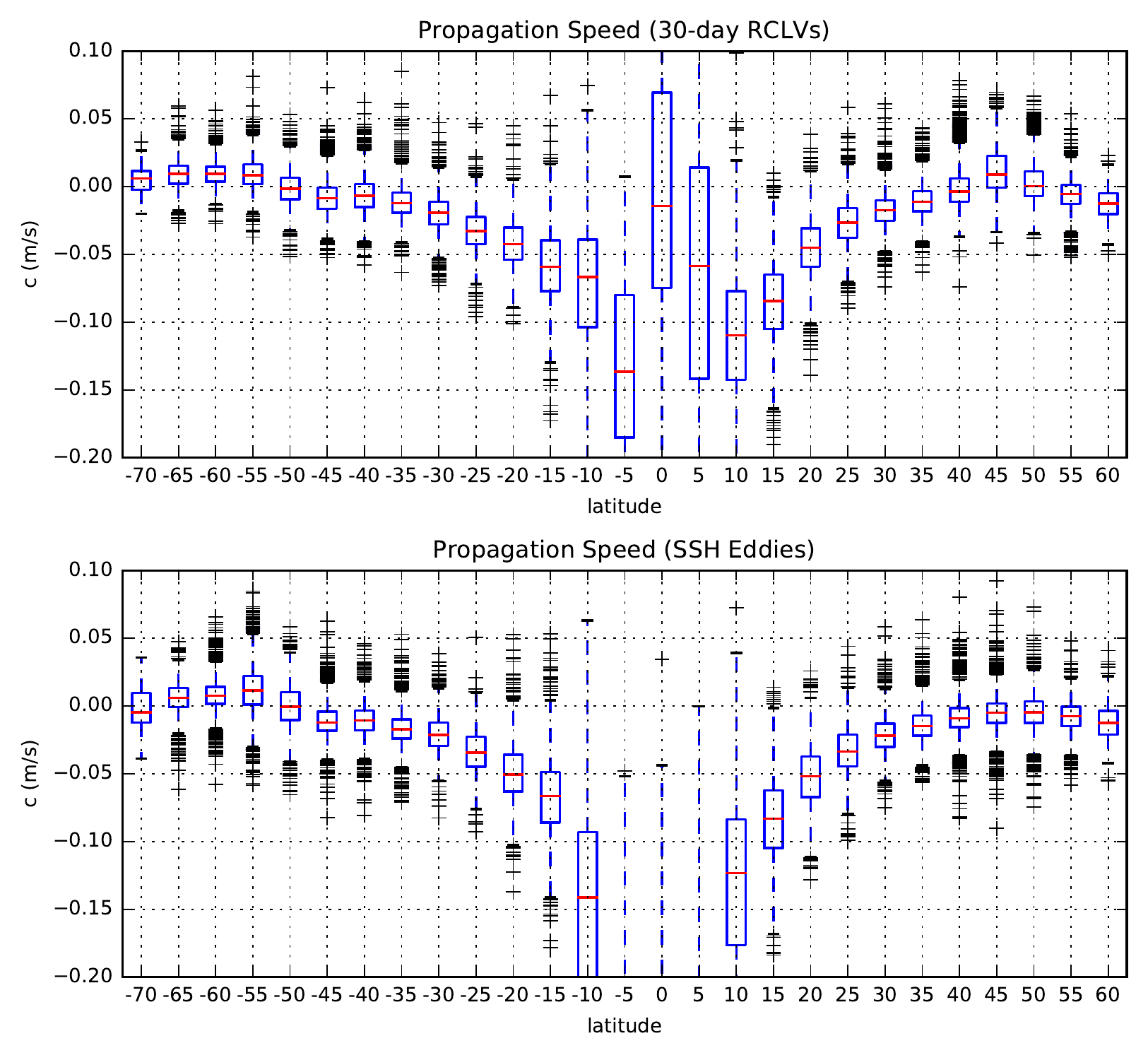}}
  \caption{As in Fig.~\ref{fig:radius}, but for eddy zonal propagation speed.
  }\label{fig:phase_speed}
\end{figure}

In Fig.~\ref{fig:phase_speed}, we examine the zonal propagation speed of RCLVs and SSH eddies.
In both cases, the zonal propagation speed is calculated as the total zonal distance traveled over the eddy lifetime.
From this point of view, RCLVs and SSH eddies look very similar, with much faster propagation at low latitudes due to the larger gradient in Coriolis parameter \citep[i.e.~$\beta$-effect; ][]{CheltonEtAl2011,KlockerAbernathey2014,AbernatheyWortham2015}.
Again, the statistics diverge in the topics, where there are vanishingly few RCLVs.

Since the RCLVs and SSH eddies propagate at the same speed, one might expect that the Eulerian methods of CSS11 and the Lagrangian method used here identify broadly similar structures.
However, the radius statistics suggest that the rotationally coherent core of mesoscale eddies is considerably smaller (by half) than the radius inferred from by CSS11. 
Furthermore, there is a significant difference in eddy lifetime; the SSH eddies last much longer than the RCLVs.
This might indicate that many SSH eddies represent materially leaky dynamical structures, which exchange water with their surroundings.
While we have not conducted a comprehensive investigation of RCLV lifetime, the fact that there were essentially zero 270-day RCLVs in the sector provides an upper bound for the timescale of this leaky exchange.
On the other hand, \citet{WangEtAl2015} found numerous materially coherent Agulhas eddies with 360-day lifetimes, suggesting that different regions of the ocean may generate less leaky eddies.
This discrepancy may be addressed in future work by applying the RCLV method at a global scale.

\section{Meridional Transport by Lagrangian Vortices}

Having described the method of identifying of coherent Lagrangian vortices, we now turn to the central question of our study: the role of these RCLVs in material transport.
For each 30 and 90-day interval, we compute the absolute diffusivity $K$ (eq.~\ref{eq:Kabs}) and relative diffusivity $K_{rel}$ (eq.~\ref{eq:Krel}) as a function of latitude using the full ensemble of Lagrangian particles. We also compute the fractional relative diffusivity $K_{rel}^{cs}$ \eqref{eq:Krel_frac}, using only particles inside the RCLVs.
As discussed in Sec.~2, $K_{rel}^{cs}$ represents the diffusivty which would result if water parcels outside of the RCLVs moved only with the zonal mean flow and induced no relative dispersion.

\begin{figure}[h]
 \centerline{\includegraphics[width=\textwidth]{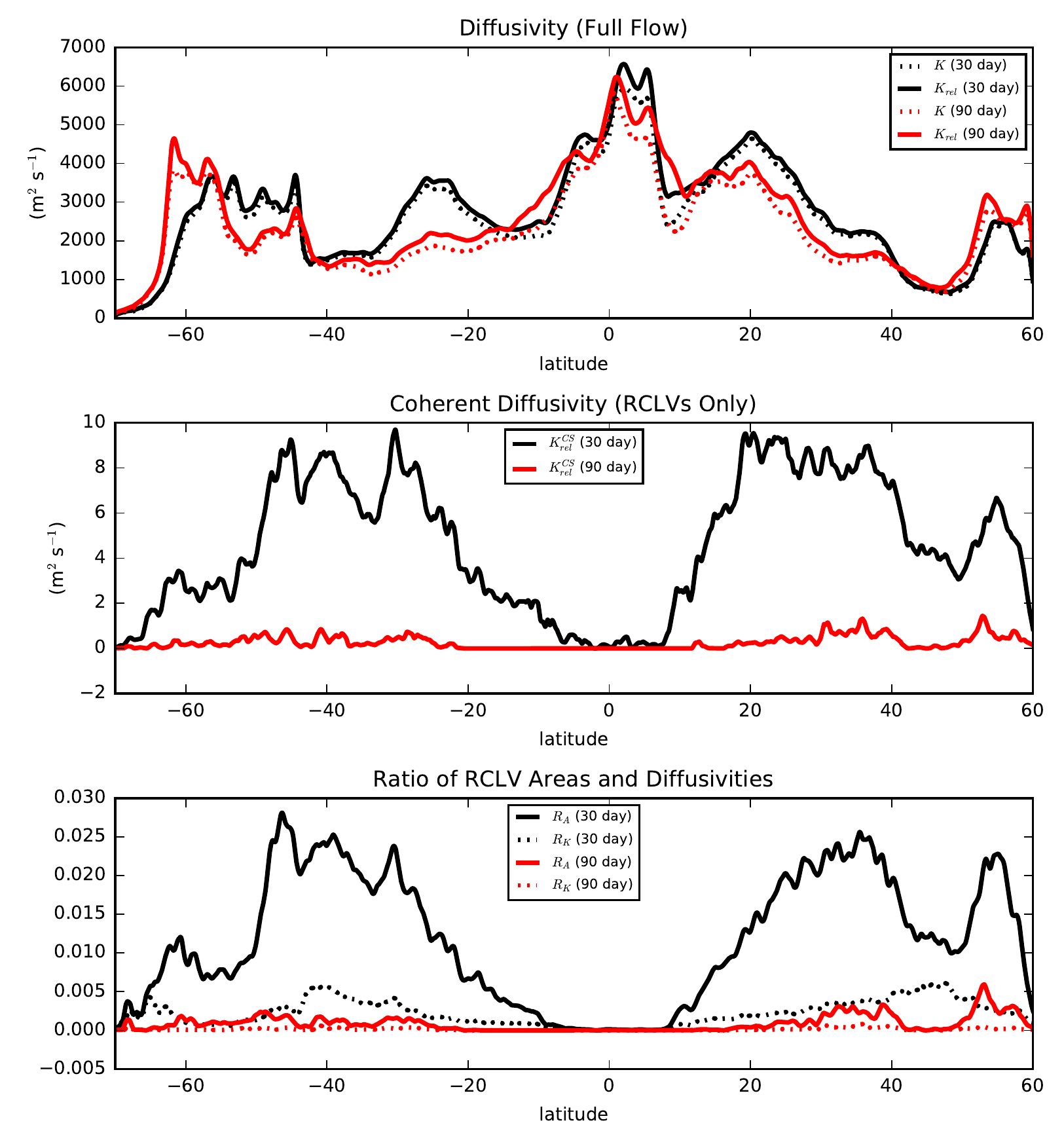}}
  \caption{Upper panel: full relative diffusivity $K_{rel}$ \eqref{eq:Krel} averaged for all 30- (blue) and 90-day (green) intervals. Middle panel: fractional relatively diffusivity $K_{rel}^{cs}$ due only to the motion of RCLVs. Note the different scale on the y-axis. Bottom panel: non-dimensional ratios of RCLV area ($R_A$) and diffusivity ($R_K$) to total area and diffusivity.
  }\label{fig:diffusivity}
\end{figure}

The results of these diffusivity calculations are shown in Fig.~\ref{fig:diffusivity}.
First, we note that there is minimal difference between $K$ and $K_{rel}$, revealing that there is negligible mean meridional advection throughout the sector.
$K_{rel}$ ranges from 500 - 6000 $\mms$, with highest values found in the tropics.
This is broadly consistent with previous estimates from this sector \citep{ZhurbasOh2003,AbernatheyMarshall2013,KlockerAbernathey2014,AbernatheyWortham2015}.
Precise agreement with previous studies is not necessarily expected, since relative diffusivity depends sensitively on the time interval \citep{Okubo1971,OllitraultEtAl2005,Lacasce2008}.
For homogeneous flows, convergence is expected for long time scales \citep{KlockerEtAl2012b}, but the diffusivities here correspond precisely to 30- and 90-day time intervals.
The difference between the 30- and 90-day results show that convergence has not been reached everywhere.

The emphasis here is not the precise value of $K_{rel}$ but rather the comparison with $K_{rel}^{cs}$ (middle panel).
The most striking difference is the order of magnitude: $K_{rel}^{cs}$ does not exceed 10 $\mms$ for 30-day RCLVs
and does not exceed 1 $\mms$ for 90-day RCLVs.
We can quantify the fraction of transport accomplished by RCLVs at each latitude via the ratio $R_K = K_{rel}^{cs} / K_{rel}$, as plotted in in Fig.~\ref{fig:diffusivity} (bottom panel).
This fraction would be close to one if most of the transport were by coherent vortices; instead, we observe that it never exceeds 0.005 for 30-day RCLVS and is an order of magnitude smaller for 90-day RCLVS.
This small contribution of RCLVs to the meridional transport mirrors the finding of \citet{WangEtAl2015} that materially coherent Agulhas eddies make a very small contribution to net transport in that region---we return to this point in the discussion.

The similarity between the shape of $K_{rel}^{cs}$ and the average density of RCVLs (Fig.~\ref{fig:eddy_count}) suggests that the primary control on $K_{rel}^{cs}$ is simply the density of RCLVs found at a particular latitude.
To test this hypothesis, we compute the RCLV area fraction $R_A$ (Fig.~\ref{fig:diffusivity}, bottom panel), which represents the average fraction of the ocean surface area that lies within an RCLV in each latitude band.
For 30-day RCLVs, $R_A$ peaks at around 0.025 and is an order of magnitude smaller for 90-day RCLVs.
This is significantly higher than $R_K$, the diffusivity fraction, revealing that the RCLVs are actually regions of anomalously {\em low} meridional dispersion.
This is unsurprising, since the RCLVs remain coherent by construction and undergo relatively low filamentation compared to the background flow.
In other words, randomly selected patches of ocean with the same surface area as the identified RCLVs would experience stronger meridional diffusion that the actual RCLVs.

\section{Discussion and Conclusion}

Many prior studies have attempted to track mesoscale eddies by following anomalies in the SSH field (and associated instantaneous surface geostrophic velocity field) through time \citep[e.g.][]{CheltonEtAl2011,DongEtAl2011,FaghmousEtAl2015}.
While such tracked Eulerian eddies may be useful for some applications, the work of \citet{BeronVeraEtAl2013} and \citet{HallerBeronVera2013} has shown that structures identified in this way are not generally materially coherent: significant material leakage can occur through the supposed structure boundaries.
This finding calls into question studies such as those of \citet{DongEtAl2014} and \citet{ZhangEtAl2014}, who attempt to infer heat, salt, and mass transports based on the displacement of tracked Eulerian eddies.
The goal of our study was to examine the material transport of mesoscale eddies defined as Lagrangian Coherent Structures.

We identified coherent eddies across a broad sector in the Eastern Pacific using an objective, Lagrangian method based on the vorticity, the so-called Rotationally Coherent Lagrangian Vortex (RCLV) approach of \citet{HallerEtAl2016}.
This computationally demanding task required the numerical advection of millions of virtual Lagrangian particles over a period of 25 years, the length of the satellite altimetry record.
To our knowledge, our study is the largest-scale application of objective Lagrangian eddy detection to date.
This comprehensive census of RCLVs in the sector allowed us to a) calculate some statistical properties of RCLVs and b) compute their contribution to net meridional transport, via a the coherent relative diffusivity $K_{rel}^{cs}$.  

The occurrence frequency, length scales, and propagation speeds of RCLVs in this sector were found to be qualitatively similar to those of SSH eddies identified by \citet{CheltonEtAl2011}.
RCLVs were larger at low latitude, consistent with the meridional variations in the baroclinic Rossby deformation radius; the RCLV radii, however, were smaller than the SSH eddy radii by about a factor of two.
A more striking difference was the eddy lifetime; while we didn't systematically examine the dependence on the extraction interval, we found essentially no RCLVs with lifetimes longer than 270 days.
We suggested that this sets an upper bound on the leakiness timescale of mesoscale eddies in this sector.
The comparison between RCLVs and SSH eddies raises many further questions about the relationship between the different methods.
How often are RCLVs embedded inside SSH eddies?
Is it possible to quantify the leakiness of the SSH eddies?
Is the Eulerian nonlinearity parameter $U/c$ (with $u$ the azimuthal eddy velocity and $c$ the translation speed) of \citet{CheltonEtAl2011} related to the presence of RCLVs?
These questions are ripe for exploration in future work.

Our primary focus here has been the calculation of meridional transport by RCLVs.
Our key finding is that $K_{rel}^{cs}$, representing the diffusive meridional transport due to RCLVs, is hundreds of times smaller than $K_{rel}$, the diffusive meridional transport of the full flow.
This means means that transport by RCLVs makes a negligible contribution to the net meridional dispersion.
By process of elimination, we can then conclude that, in this sector, meridional dispersion is primarily by {\em incoherent} motions, outside of the RCLV boundaries.
This conclusion is in contradiction with the claims of \citet{DongEtAl2014} and \citet{ZhangEtAl2014}, who base their transport estimates on Eulerian eddy tracking.
Our conclusion {\em is}, however, highly consistent with the findings of \citet{WangEtAl2015}, who used a different objective Lagrangian eddy identification method to quantify material transport by Agulhas rings.
Indeed, \citet{WangEtAl2015} found that the cross-Atlantic transport by materially coherent Agulhas eddies was two orders of magnitude smaller than prior estimates based on Eulerian eddy tracking.
Since Agulhas rings move only in one direction, \citet{WangEtAl2015} described the transport in terms of advection.
In our sector, where eddies drift both north and south with equal frequency, transport was quantified in terms of diffusion.
(A more general statistical decomposition of transport by coherent Lagrangian eddies would involve both advection and diffusion.)
Regardless, the overall conceptual agreement between their study and ours, which used a different method and examined a different region, suggests that relatively small material transport by coherent Lagrangian eddies is a robust result.

This finding does not mean, however, that coherent mesoscale eddies are insignificant for transport.
Indeed, studies in the spectral domain \citep[e.g.][]{KillworthEtAl2004,AbernatheyWortham2015} show that the eddy flux peaks at length scales and phase speeds associated with mesoscale eddies.
Our results here, however, suggest that the meridional eddy transport is likely driven by stirring and filamentation on the periphery of coherent eddies, rather than by coherent meridional motion of the eddy core. 
This mechanism was illustrated clearly by \citet{HausmannCzaja2012}, who studied eddy heat transport by examining the cross-correlation structure between satellite-observed SSH and SST anomalies.
Using the cross-correlation, they decomposed the Eulerian eddy heat flux into a drift component (associated with translation of fluid within the eddy core) and a swirl component, associated with peripheral stirring.
They found that the swirl component was large enough to make a leading-order contribution to the oceanic heat budget, but that the drift component was negligible.
This finding is compatible with our results, but not with \citet{DongEtAl2014}, who reached the opposite conclusion through methods similar to \citet{HausmannCzaja2012}.

The chief limitation of the results we have presented is  their reliance on the AVISO surface geostrophic velocities, which we have simply accepted at face value as adequately representative of the near-surface flow.
In reality, there are many potential sources of error in these velocity fields including measurement error of the altimeter itself, limited spatial and temporal sampling, mapping errors related to the gridding of satellite tracks, and the presence of ageostrophic and vertical velocities.
Lagrangian coherent structures represent stable attractors of the flow and are robust to the presence of small noise \citep{Haller2015}.
The spatial and temporal sampling issue, however, is likely more serious; \citet{KeatingEtAl2012} showed that such subsampling can seriously degrade the finite-time Lyapunov exponent field in idealized turbulence simulations.
It is an open question how the presence of submesoscale flows and internal waves impacts the leakiness of mesoscale transport barriers and eddies.
A comprehensive investigation of the observational errors in the detection of Lagrangian coherent structures from satellite altimetry observations would indeed be a valuable contribution.
The full three-dimensional structure of RCLVs also remains an open question which is not possibly to address using satellite observations alone.
Analysis of a high-resolution general circulation model would be a good way to probe this question.

We see this study as the first step towards a fully global characterization of mesoscale coherent structures.
It is our hope that this sort of detailed description of the Lagrangian kinematics of mesoscale transport will eventually lead to more accurate parameterization of mesoscale transport in coarse-resolution climate models.
In the case of the Eastern Pacific, it appears that we can reliably neglect long-range meridional transport due to fluid trapping within coherent eddy cores.
This is good news from the perspective of parameterization, since the swirling mode of eddy transport seems more amenable to representation via diffusive closures.
Determining whether such a conclusion holds more generally will have to await the completion of a global-scale Lagrangian eddy census, which is a serious computational challenge.

%



\acknowledgments
R.P.A.~gratefully acknowledges the support of an NSF CAREER award (OCE 1553593).
We thank Nathaniel Tarshish, Ivy Frenger, Carolina DuFour, and Steve Griffies for their feedback on this manuscript.

 \bibliographystyle{ametsoc2014}
 \bibliography{references}

%
%


\end{document}